\documentclass{article}\begin{document}\centerline{Analysis of the second order exchange self energy of a dense electron gas}\vskip .8in\centerline{M.L. Glasser$^{1}$ and George Lamb$^{2}$}

\centerline{$^{1}$Center for Quantum Device Technology and Department of Physics, Clarkson University}\centerline{Potsdam, NY 13699-5820, USA}
\centerline{$^{2}$2942 Ave. del Conquistador, Tucson, AZ 85749-9304, USA}
\vskip 2in\centerline{\bf Abstract}\vskip .1in
\noindent
We investigate the evaluation of the six-fold integral representation for the second order exchange contribution to the self energy of a dense three dimensional gas on the Fermi surface.

\vskip 1in
\noindent PACS 71.10.CA, 05.30.Fk
\newpage
\centerline{\bf Introduction}\vskip .1in

The second order exchange energy, represented by the diagram in Fig.1a contributes importantly to the correlation energy of a dense electron gas [1].  It is given by the nine-fold integral
$$E_{2x}=\frac{3}{32\pi^4}\int d^3p_1\int d^3p_2\int \frac{dq^3}{q^2}\frac{f_{p_1}f_{p_2}f'_{p_1+q}f'_{p_2+q}}{(\vec{q}+\vec{p}_1+\vec{p}_2)^2(q^2+\vec{p_1\cdot  \vec{q}+\vec{p}_2\cdot\vec{q})}}\eqno(1)$$
in three dimensions, where $f_p$ denotes the Fermi distribution function for electrons of wave vector $\vec{p}$ and $f'_{p}$ denotes that for holes. In a remarkable display of mathematical virtuosity (1) was evaluated in closed form by Onsager[2] and Onsager, Mittag and Stephen[3] who found
$$E_{2x}=\frac{1}{6}\ln(2)-\frac{3}{4\pi^2}\zeta(3).\eqno(2)$$
Subsequently, Ishihara and Ioratti[4] worked out the corresponding value for a two-dimensional system, and the d-dimensional case was evaluated by Glasser[5].

Recently the the second order exchange term in the electron self energy, represented by the diagram in Fig.1b was studied by Ziesche[6]. It is given, in three dimensions, by the six-fold integral
$$\Sigma_{2x}(k)=\frac{1}{4\pi^4}\int\frac{d^3q}{q^2}\int d^3p\frac{f_pf_{k+q}f_{p+q}f'_pf'_{p+q}}{(\vec{k}+\vec{p}+\vec{q})^2(q^2+\vec{k}\cdot\vec{q}+\vec{p}\cdot\vec{q})}.\eqno(3)$$
For $k=k_F(=1)$ Ziesche succeeded in decomposing (3) into the sum $\Sigma_{2x}=-(X_1+X_2)/4\pi^2$ of the two simpler integrals
$$X_1=\int\frac{d^3q_1}{q_1^2}\int \frac{d^3q_2}{q_2^2}\frac{f_{k+q_1+q_2}f'_{k+q_1}f'_{k+q_2}}{\vec{q}_1\cdot\vec{q}_2}$$
$$X_2=-\int\frac{d^3q_1}{q_1^2}\int\frac{ d^3q_2}{q_2^2}\frac{f'_{k+q_1+q_2}f_{k+q_1}f_{k+q_2}}{\vec{q}_1\cdot\vec{q}_2}\eqno(4)$$
and by following the procedure in [3], he managed to perform three of the integrations, thereby obtaining
$$X_1=-16\pi\int_0^1dp\int_0^1dq\int_{-1}^1\frac{dx}{(1-p^2q^2)}\frac{F[p,q,x]}{1+q^2}$$
$$X_2=16\pi\int_0^1dp\int_0^1dq\int_{-1}^1\frac{dx}{(1-p^2q^2)}\frac{q^2F[p,q,x]}{1+q^2}\eqno(5)$$
where
$$\alpha=\frac{1-q^2}{2q},\mbox{\hskip .3in}\beta=\frac{1-p^2}{2p},\mbox{\hskip .3in}a=\frac{1+p^2q^2}{2pq}$$
$$F[p,q,x]=\frac{2}{a^2-x^2}\tan^{-1}\left[\frac{\alpha x+\beta}{\sqrt{(1+\alpha^2)(1-x^2)}}\right].\eqno(6)$$
The integrals in (6) are suitable for numerical evaluation and Ziesche found $X_1=-30.70598\dots$,$X_2=21.28490\dots$.

According to the Hugenholtz-van Hove- Luttinger-Ward theorem[7] $\Sigma_{2x}=E_{2x}$, so
$$X_1+X_2=3\zeta(3)-\frac{2\pi^2}{3}\ln(2).\eqno(7)$$
The aim of this note is to evaluate $X=X_2-X_1$ analytically, so as to obtain closed form expressions for the integrals in (4).
\vskip .2in

\centerline{\bf Calculation}\vskip .1in

From (5) we have
$$X=16\pi\int_0^1dq\int_0^1dp\int_{-1}^1dx\frac{F[p,q,x]}{1-p^2q^2}.\eqno(8)$$
Since the limits on the $x$-integral are symmetric, we retain only the even part of the integrand of (8)
by averaging $X$ and the integral obtained by $x\rightarrow-x$ and combining the two arctangents, thus obtaining
$$X=16\pi\int_0^1dp\int_0^1dq\int_0^1dx\frac{\tan^{-1}\left[\frac{2\beta\sqrt{(1+\alpha^2)(1-x^2)}}{\alpha^2-\beta^2+1-x^2}\right]}{(1-p^2q^2)(a^2-x^2)}.\eqno(9)$$
Next, we set $q=e^{-u}$, $p=e^{-v}$, $x=\sin\phi$, so $\alpha=\sinh\; u$, $\beta=\sinh\; v$, $a=\cosh(u+v)$, and
$$X=$$
$$8\pi\int_0^{\infty}du\int_0^{\infty}dv\int_0^{\pi/2}d\phi\; \cos\;\phi\frac{tan^{-1}\left[\frac{(\sinh(u+v)+\sinh(v-u))\cos\;\phi}{\sinh(u+v)\sinh(u-v)+\cos^2\phi}\right]}{\sinh(u+v)[\sinh^2(u+v)+\cos^2\phi]}.\eqno(10)$$

We make the coordinate transformation $r=v+u$, $s=v-u$, having Jacobian 1/2, to obtain
$$X=4\pi\int_0^{\infty}dr\int_{-r}^rds\int_0^{\pi/2}d\phi\;\cos\phi\frac{\tan^{-1}\left[\frac{(\sinh\; r+\sinh\; s)\cos\; \phi}{\cos^2\phi-\sinh\; r\sinh\; s}\right]}{\sinh\; r(\sinh^2r+\cos^2\phi)}.\eqno(11)$$

Since
$$\tan^{-1}\left[\frac{\cos\;\phi(\sinh\; r+\sinh\; s)}{\cos^2\phi-\sinh\; r\sinh\; s}\right]=$$
$$Im\; \ln[(\cos\;\phi+i\sinh\; r)(\cos\;\phi+i\sinh\; s)]=$$
$$\tan^{-1}\left(\frac{\sinh\; r}{\cos\;\phi}\right)+\tan^{-1}\left(\frac{\sinh\; s}{\cos\;\phi}\right),\eqno(12)$$
(11) becomes
$$X=$$
$$4\pi\int_0^{\infty}dr\int_{-r}^rds\int_0^{\pi/2}d\phi\; \cos\;\phi\frac{\tan^{-1}(\sec\;\phi\sinh\; r)+\tan^{-1}(\sec\;\phi\sinh\; s)}{\sinh\; r(\cos^2\phi+\sinh^2r)}\eqno(13)$$
Once again, we may drop the term in the integrand of (13) odd in $s$ and perform the elementary $s-$ integration, so that
$$X=8\pi\int_0^{\infty}\frac{rdr}{\sinh\; r}\int_0^{\pi/2}d\phi\; \tan^{-1}\left(\frac{\sinh\; r}{\cos\;\phi}\right)\frac{\cos\;\phi}{\cos^2\phi+\sinh^2r}.\eqno(14)$$
To evaluate the $\phi-$integral, we set $\tan\;\psi=\sec\;\phi\sinh\; r$ , $\mu=\tan^{-1}(\sinh\; r)=\cos^{-1}(sech\; r)$, to transform (14) into
$$X=8\pi\int_0^{\infty}\frac{rdr}{\sinh\; r}\cos\;\mu\int_{\mu}^{\pi/2}\frac{\psi\cos\;\psi\; d\psi}{\sqrt{\sin^2\psi-\sin^2\mu}}.\eqno(15)$$
The $\psi-$ integral is tabulated[8] and $X$ is reduced to a single integral
$$X=4\pi^2\int_0^{\infty}\frac{r\; sech\; r\ln(1+sech\; r)}{\sinh\; r}dr.\eqno(16)$$

To evaluate the remaining integral, let 
$$f(a)=\int_0^{\infty}\frac{r\ln(1-a\; sech\; r)}{\sinh\; r\cosh\; r}dr\eqno(17)$$
for which $f(1)=X/4\pi^2$ and $f(0)=0$.  By differentiation with respect to $a$ and partial fraction decomposition, we obtain $$(1-a^2)\frac{df}{da}=$$
$$\int_0^{\infty}\frac{rdr}{\sinh\; r}-2a\int_0^{\infty}\frac{rdr}{\sinh\; 2r}-\frac{1}{a}\int_0^{\infty}r\sinh\; r\left[\frac{1}{\cosh\; r}-\frac{1}{\cosh\; r+a}\right].\eqno(18)$$
The first two integrals on the right hand side of (18) are tabulated[9] and, after an integration by parts, we find
$$(1-a^2)\frac{df}{da}=\frac{\pi^2}{8}(2-a)-\frac{1}{a}\int_0^{\infty}\ln(1+a \; sech\; r)dr\eqno(19)$$
The substitution $u=sech\; r$ leads to another tabulated integral[10], giving
$$\frac{df}{da}=-\frac{\pi^2}{8a}\left(\frac{1-a}{1+a}\right)+\frac{1}{2a}\frac{(\cos^{-1}a)^2}{1-a^2},\eqno(20)$$
which, with the substitution $a=\cos\;\theta$ yields
$$X=4\pi^2\int_0^1\frac{df}{da}da=\pi^4\ln(2)+4\pi^2\int_0^{\pi/2}\frac{d\theta}{\sin\;2\theta}[\theta^2-\frac{\pi^2}{8}(1-\cos(2\theta))].\eqno(21)$$
Finally, we find by setting $\phi=2\theta$,  and folding the new range of integration $[\pi/2,\pi]$ back to $[0,\pi/2]$
$$X=\pi^4\ln(2)+4\pi^2\int_0^{\pi/2}\frac{4\phi(\phi-\pi)}{\sin\;\phi}d\phi=$$
$$\pi^4\ln(2)-\frac{7}{2}\pi^2\zeta(3),\eqno(22)$$
where we have used[11]
$$\int_0^{\pi/2}\frac{\phi d\phi}{\sin\;\phi}=2{\bf{G}},\mbox{\hskip .4in}\int_0^{\pi/2}\frac{\phi^2d\phi}{\sin\;\phi}=2\pi{\bf{G}}-\frac{7}{2}\zeta(3)\eqno(23)$$
in which ${\bf{G}}$ denotes Catalan's constant.

\vskip .2in\centerline{\bf Discussion}\vskip .1in

Our result is that we have obtained closed form expressions for the two six-fold integrals in (4)
$$X_1=-\pi^4\left[\frac{4}{3}\ln(2)-\frac{5}{\pi^2}\zeta(3)\right]= \eqno(24)$$
$$-30.70598523924889925762268444608481536875855208165945918981645846\dots$$
$$X_2=\pi^4\left[\frac{2}{3}\ln(2)-\frac{2}{\pi^2}\zeta(3)\right]=\eqno(25)$$
$$21.284905670516337983402598547497784400625730440810132220995696061\dots$$

This gives the value
$$\Sigma_{2x}=\eqno(26)$$
$$0.0241791589181444058954507621628984314049152384251207335945309986\dots$$
in agreement with Ziesche's [6] seven place calculation. We hope to extend our calculation to an electron gas of arbitrary dimension, as was done for $E_{2x}$.

\vskip .2in\noindent
{\bf Acknowlegements}\vskip .1in

The first author thanks Dr. Paul Ziesche for a discussion of his work and the National Science foundation for support under Grant DMR 0121146.
\newpage
\noindent
{\bf{References}}
\vskip .1in\noindent
[1] M. Gell-Mann and K. Brueckner, Phys. Rev. {\bf{106}}, 364 (1957).

\noindent
[2] L.Onsager, Unpublished.

\noindent
[3]  L. Onsager, L. Mittag and M.J. Stephen, Ann. Physik (Leipzig) {\bf{18}}, 71 

(1966).

\noindent
[4]   A. Isihara and L. Ioriatti, Phys. Rev. {\bf{B22}}, 214 (1980).

\noindent
[5] M.L. Glasser, J. Comp. Appl.Math.{\bf{10}}, 293 (1984).

\noindent
[6]  P. Ziesche, ArXiv.org cond-mat/0605188. (Annalen der Physik, In Press.)

\noindent
[7]  J.M. Luttinger and J.C. Ward, Phys. Rev.{\bf{118}}, 1417 (1960).

\noindent
[8] I.S. Gradshteyn and I.M. Ryzhik, {\it Table of Integrals, Series and Products}, 

6th ed., Academic Press(2000). p. 466, No. 3.842(2).

\noindent
[9] Ref.8, p.369, No.3.521(1)

\noindent
[10] Ref. 8, p.554, No.4.292(5).

\noindent
[11] Ref. 8, p. 427, Nos. 3.747(1,2).

\end{document}